# High-resolution soft x-ray photoemission study of a Kondo semiconductor and related compounds

A. Sekiyama[a], Y. Fujita[a], M. Tsunekawa[a], S. Kasai[a], A. Shigemoto[a], S. Imada[a], D. T. Adroja[b], T. Yoshino[b], F. Iga[b], T. Takabatake[b], T. Nanba[c], and S. Suga[a]

[a]Department of Material Physics, Graduate School of Engineering Science, Osaka University, Toyonaka, Osaka 560-8531, Japan
[b]ADSM, Hiroshima University, Higashi-Hiroshima 739-8530, Japan
[c]Department of Physics, Kobe University, Kobe 657-8501, Japan

**Abstract**

We have performed the bulk-sensitive high-resolution soft x-ray photoemission study of a Kondo semiconductor CeRhAs and related compounds CeNiSn and CePdSn. The comparison of the spectra of polycrystalline CePdSn on the fractured and scraped surfaces shows that the fracturing of the samples is much better than the scraping in order to obtain intrinsic photoemission spectra. The Ce 4d core-level spectra show clear differences in the electronic states among the materials.

*Key words: Photoemission, CeRhAs, CePdSn, strongly correlated systems*

## 1. Introduction

The formation of the (pseudo-) gap in Kondo semiconductors, which are nonmagnetically semiconducting at low temperatures and behave as local-moment systems at high temperatures, is one of the intriguing phenomena in the strongly correlated systems. CeRhAs is known as the Kondo semiconductor with the Kondo temperature of the order of 1000 K, namely, a strongly hybridized 4f system [1,2]. Although CeNiSn and CePdSn are isostructual with CeRhAs, the hybridization between the 4f electrons and valence-bands is weaker than that in CeRhAs where the Kondo temperature is ~50 and < 7 K, respectively [3,4]. Low-energy photoemission (PES) studies have so far been performed on the scraped [5] and fractured [2] surfaces of CeRhAs in order to investigate its electronic structures. On the other hand, it is known that the low-energy PES often probes the surface electronic states strongly deviated from the bulk states [6,7]. In this paper, we show the bulk-sensitive high-energy PES spectra of CeRhAs, CeNiSn and CePdSn. We have also examined the differences in the spectra on the scraped and fractured surfaces of CePdSn.

## 2. Experimental

The PES measurements were performed at BL25SU in SPring-8. The base pressure was about $4 \times 10^{-8}$ Pa. In order to obtain clean surfaces, the samples were fractured or repeatedly scraped *in situ* as discussed later at the measuring temperature of 20 K. The surface cleanliness was confirmed by the absence or weakness of the possible O 1s and C 1s signal. The binding energies in the spectra were calibrated by the measurement of the gold electrically connected to the samples.

## 3. PES spectra of CePdSn

In order to know which surface (fractured or scraped) is better to obtain the intrinsic photoemission spectra even at the high-energy soft x-ray excitation, we have measured the several core-level PES spectra of *polycrystalline* CePdSn on both fractured and scraped surfaces. As shown in Fig. 1, all core-level peaks in the spectra on the scraped surface are obviously broader than those on the fractured surface. The gravity of the spectral weight on the fractured surface is shifted to the lower binding energy side compared with that on the scraped surface. The same tendency has also observed for the Sn $3d_{3/2}$ and Pd $3d_{3/2}$ peaks as mentioned below. The full width of half maximum (FWHM) of the Sn $3d_{5/2}$ and Pd $3d_{5/2}$ peaks in the spectra on the fractured surfaces are estimated as about 630 and 420 meV while those on the scraped surface are about 850 and 620 meV. In the case of an intermetallic compound CePdSn, any complicated structure such as final-state multiplet structure is not expected near the main peaks in the Sn 3d, 4d and Pd 3d spectra. Indeed, it seems that there is not any shoulder or pre-peak in the Sn $3d_{5/2}$ as well as Pd $3d_{5/2}$ spectra on the scraped surface while a shoulder seems to be seen at the higher binding energy side (by ~0.3 eV) of the main peak in the Sn 4d spectra on the scraped surface. Although an origin of the broader FWHM of the peak in the spectra on the scraped surface is not clear at present, it is natural to judge that the fractured surface is better than the scraped surface in order to obtain the intrinsic spectra.

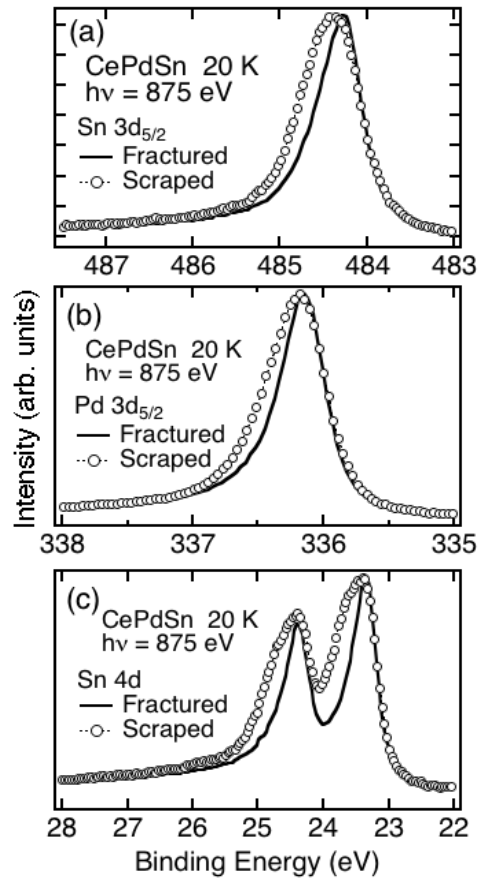

Fig. 1. Comparison of the core-level photoemission spectra of CePdSn on the fractured and scraped surfaces. (a) Sn $3d_{5/2}$ core level. (b) Pd $3d_{5/2}$ core level. (c) Sn 4d core level. The energy resolution for these spectra were set to 200 meV.

Figure 2 shows the valence-band PES spectra near the Fermi level ($E_F$) of polycrystalline CePdSn with the resolution of ~100 meV measured at hν = 875 eV, which is below the Ce 3d absorption threshold. Therefore, the spectral weight is mainly due to the Pd 4d and Sn 5p states. In both spectra are seen the clear Fermi cut-off. There are broad peaks at ~0.2 and ~1.0 eV in the spectrum on the fractured surface. We consider that these originate from the Pd 4d-Sn 5p antibonding states reflecting the band structures of CePdSn [8]. On the other hand, these structures are somewhat smeared in the spectrum on the scraped surface. We consider that such a phenomenon is caused by the formation of the scraped surface region, in which the translational symmetry of the crystal is rather broken, comparable or longer than the probing depth of the high-energy soft x-ray PES (at most 20 Å when the photoelectron kinetic energy of <~1000 eV). Possible charge up of the scraped surface might be the other origin of the broadened peak structures near $E_F$.

The results seen in Figs. 1 and 2 indicate that the fractured surfaces are better to obtain the intrinsic PES spectra if we would like to discuss the electronic states of solids based on the PES spectra within an energy scale of several hundreds meV.

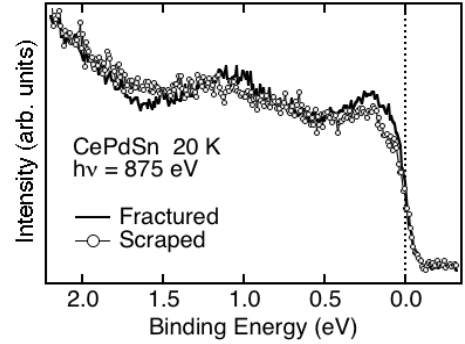

Fig. 2. High-resolution (~100 meV) soft x-ray photoemission spectra of CePdSn on the fractured and scraped surfaces.

## 4. Ce 4d core-level spectra

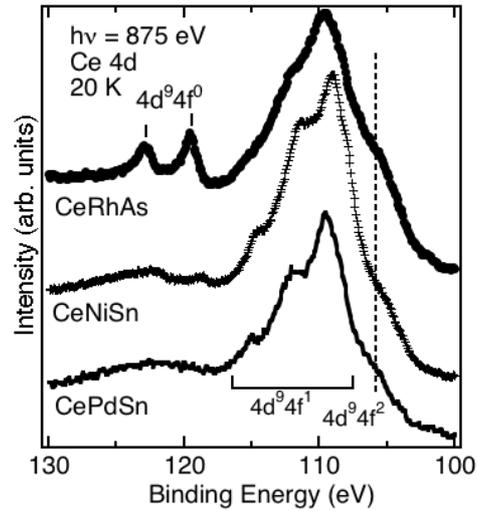

Fig. 3. Ce 4d core-level photoemission spectra of CeRhAs, CeNiSn and CePdSn on the fractured surfaces. The energy resolution for these spectra were set to 350 meV.

Figure 3 shows the Ce 4d core-level PES spectra of CeRhAs, CeNiSn and CePdSn on the fractured surfaces. In all spectra, a strong peak structure including several peaks or shoulders is observed in an energy region from

108 to 117 eV, which corresponds to the $4d^9 4f^1$ final states. A shoulder seen at 106 eV is ascribed to the $4d^9 4f^2$ final states, whose intensity becomes stronger on going from CePdSn to CeRhAs, namely from the weakly to strongly hybridized systems. In addition, there are two peaks at 119.5 and 123 eV in the spectrum of CeRhAs. These are due to the $4d^9 4f^0$ final states, where the energy between these two peaks reflects the spin-obit splitting of the 4d orbitals [9]. The $4d^9 4f^0$ contributions are slightly seen in the spectrum of CeNiSn whereas those are hardly observed for CePdSn. It is known that the $4d^9 4f^0$ spectral weight is larger for the strongly hybridized system [9], therefore the relatively stronger $4d^9 4f^0$ intensity in the spectrum of CeRhAs indicates that the hybridization between the 4f electrons and valence-bands is much stronger in CeRhAs than the other compounds.

## Acknowledgements

We are grateful to A. Higashiya, A. Irizawa, A. Yamasaki, H. Fujiwara, T. Satonaka, K. Konoike, T. Muro and Y. Saitoh for supporting the experiments. This work was supported by a Grant-in-Aid for COE research (10CE2004) and Creative Scientific Research (15GS0213) from the Ministry of Education, Culture, Sports, Science and Technology (MEXT), Japan. The experiments were performed under the approval of the Japan Synchrotron Radiation Research Institute (2001A0130-NS-np).

## References

[1] T. Takabatake, F. Iga, T. Yoshino, Y. Echizen, K. Katoh, K. Kobayashi, M. Higa, N. Shimizu, Y. Bando, G. Nakamoto, H. Fujii, K. Izawa, T. Suzuki, T. Fujita, M. Sera, M Hiroi, K. Maezawa, S. Mock, H. v. Löhneysen, A. Brückl, K. Neumaier and K. Andres, J. Magn. Magn. Mater. **177-181**, 277 (1998).
[2] K. Shimada, K. Kobayashi, T. Narimura, P. Baltzer, H. Namatame, M. Taniguchi, T. Suemitsu, T. Sasakawa and T. Takabatake, Phys. Rev. B **66**, 155202 (2002).
[3] D. T. Adroja, S. K. Malik, B. D. Padalica and R. Vijayaraghavan, Solid State Commun. **66**, 1201 (1988).
[4] T. Takabatake, F. Teshima, H. Fujii, S. Nishigori, T. Suzuki, T. Fujita, Y. Yamaguchi, J. Sakurai and D. Jaccard, Phys. Rev. B **41**, 9607 (1990).
[5] H. Kumigashira, T. Sato, T. Yokoya, T. Takahashi, S. Yoshii and M. Kasaya, Phys. Rev. Lett. **82**, 1943 (1999).
[6] A. Sekiyama, T. Iwasaki, K. Matsuda, Y. Saitoh, Y. Onuki and S. Suga, Nature **403**, 396 (2000).
[7] A. Sekiyama, K. Kadono, K. Matsuda, T. Iwasaki, S. Ueda, S. Imada, S. Suga, R. Settai, H. Azuma, Y. Onuki and Y. Saitoh, J. Phys. Soc. Jpn. **69**, 2771 (2000).
[8] K. Takegahara and H. Harima, unpublished.
[9] J. C. Fuggle, F. U. Hillebrecht, Z. Zolnierek, R. Lässer, Ch. Freiburg, O. Gunnarsson and K. Schönhammer, Phys. Rev. B **27**, 7330 (1983).